# The role of Junk DNA

Agoni Valentina

valentina.agoni@unipv.it

# Abstract

Many efforts have been done in order to explain the role of junk DNA, but its function remain to be elucidated. In addition the GC-content variations among species still represent an enigma. Both these two mysteries can have a common explanation: we hypothesize that the role of junk DNA is to preserve the mutations probability that is intrinsically reduced in GC-poorest genomes.

~ • ~

Over 98% of the human genome is noncoding DNA [1]. Initially, a large proportion of noncoding DNA had no known biological function and was therefore sometimes referred to as "junk DNA". The term "junk DNA" was formalized in 1972 by Susumu Ohno[2]. However, some noncoding DNA is transcribed into functional non-coding RNA molecules, e.g. transfer RNA, ribosomal RNA, regulatory RNAs, some other sequences include origins of replication, centromeres and telomeres. Over 8% of the human genome is made up of (mostly decayed) endogenous retrovirus sequences, as part of the over 42% fraction that is recognizably derived of retrotransposons, while another 3% can be identified to be the remains of DNA transposons. Much of the remaining half of the genome that is currently without an explained origin is expected to have found its origin in transposable elements that were active so long ago (> 200 million years) that random mutations have rendered them unrecognizable [3]. Genome size variation in at least two kinds of plants is mostly the result of retrotransposon sequences [4][5]. Pseudogenes are dysfunctional relatives of genes that have lost their protein-coding ability or are otherwise no longer expressed in the cell [6]. Pseudogenes often result from the accumulation of multiple mutations within a gene whose product is not required for the survival of the organism. Although not protein-coding, the DNA of pseudogenes may be functional [1] similar to other kinds of non-coding DNA which can have a regulatory role. Moreover the origins and importance of spliceosomal introns comprise one of the longest-abiding mysteries of molecular evolution. Considerable debate remains over several aspects of the evolution of spliceosomal introns, including the timing of intron origin and proliferation, the mechanisms by which introns are lost and gained, and the forces that have shaped intron evolution [7].

On the other hand the evolutionary forces try to minimize genomes sizes.

There are an estimated 20,000-25,000 human protein-coding genes. The estimate of the number of human genes has been repeatedly revised down from initial predictions of 100,000 or more as genome sequence quality and gene finding methods have improved, and could continue to drop further,[8][9]. Protein-coding sequences account for only a very small fraction of the genome (approximately 1.5%), and the rest is associated with non-coding RNA molecules, regulatory DNA sequences, LINEs, SINEs, introns, and sequences for which as yet no function has been elucidated [10].

The C-value paradox claims that genome size does not correlate with organism complexity [11]. Some plants or single-celled protists have genomes much larger than that of humans. Another open question is the GC-content enigma: the genomic GC-content varies dramatically, from less than 20% to more than 70% among genomes with apparently no correlation with organism complexity.

Hildebrand et al. [12], they find a large excess of synonymous GC→AT mutations over AT→GC mutations. The GC pair is bound by three hydrogen bonds, while AT pairs are bound by two hydrogen bonds. DNA with high GC-content is more stable than DNA with low GC-content.

Both these two mysteries, the role of junk DNA and the GC-content variations, can have a common explanation: since GC→AT mutations are more probable respect to others single nucleotide substitutions, the role of junk DNA can be to preserve the mutations probability that is intrinsically reduced in GC-poorest genomes.

If we plot GC-content and genome size we found an indirect proportion between the two (Figure 1).

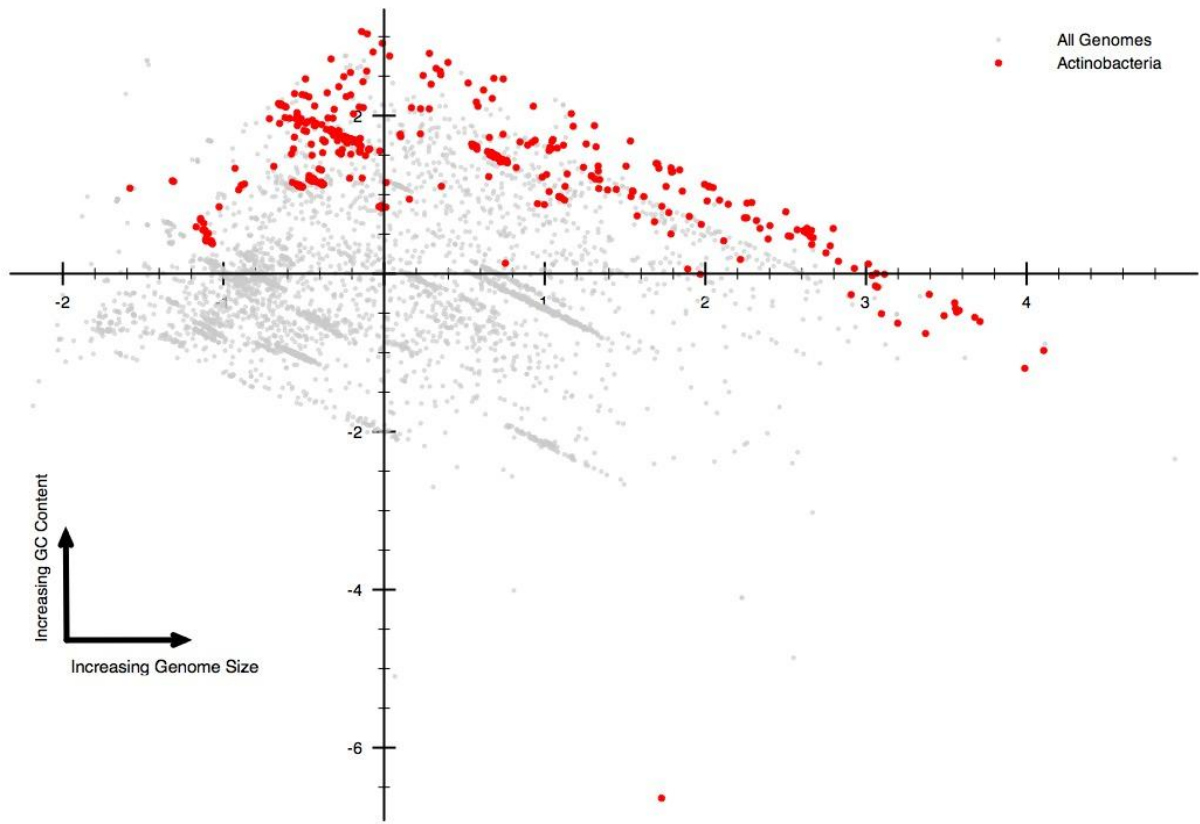

**Figure 1.** GC_Content_vs_Genome_Size_of_Different_Genomes [14].

All these considered we hypothesize that the role of junk DNA is to preserve the mutations probability that is intrinsically reduced in GC-poorest genomes.

Indeed the human genome is $3\cdot10^9$ bp long. The probability of mutation is about $10^{-8}$ per base per generation. Knowing that a protein is defined essentially by its active site (because the 3D structure of proteins derives from $\alpha$-helixes and $\beta$-sheets), we can estimate that it is defined by ~10 codons (the third codons of 10 triplets) letting down the start and stop codons. In other words we can suppose that few mutations can potentially generate a new protein in a sequence that was previously noncoding. Including both introns and exons a gene covers ~$10^4$ bp. Human genome dimensions do not allow few mutations to occur in the space covered by a CDS (coding sequence). In other words junk DNA is required for the creation of a new functional protein in few replication cycles.

On the other hand if we look at the GC-Content vs the genome size in bacteria (Figure 2) We find a direct proportion. This meaning that bacteria do not need more template to increase their mutations rate, and the actually do not have junk DNA.

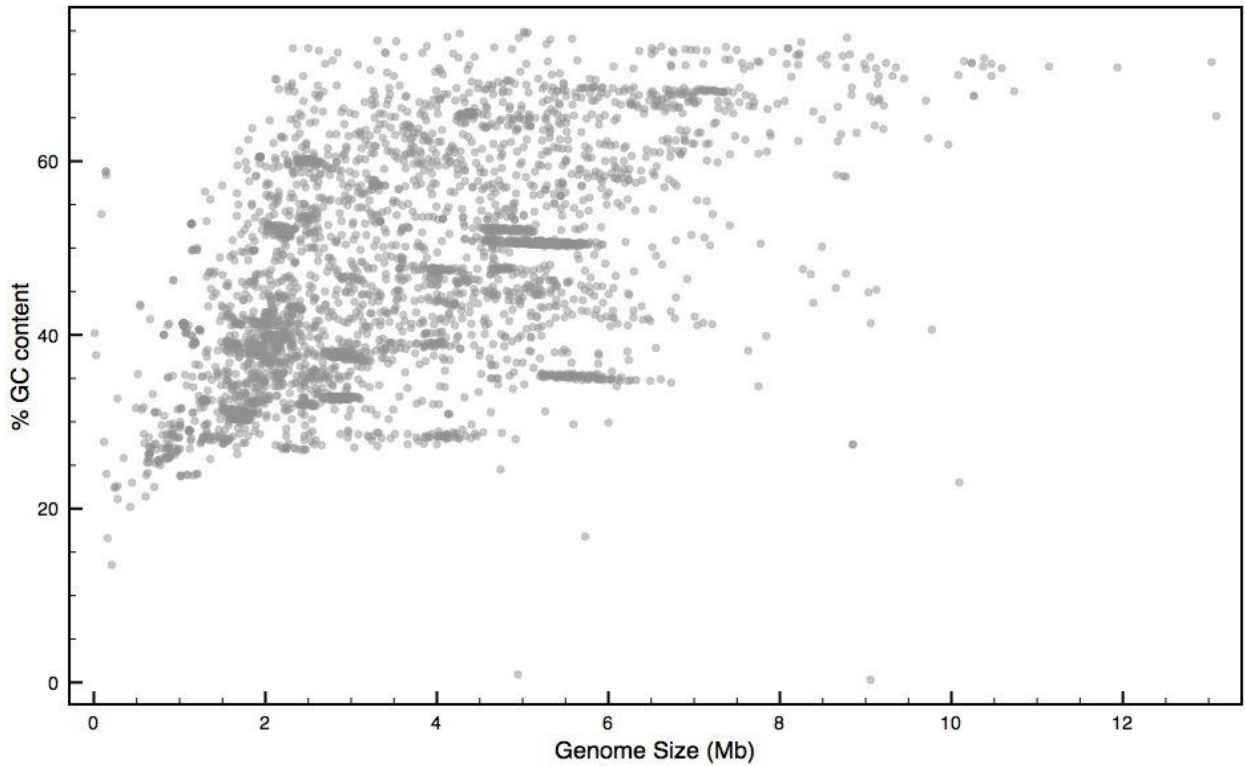

**Figure 2.** GC_Content_vs_Genome_Size_of_4231_Bacteria_Genomes [15].

This hypothesis is in accordance with the fact that only about 2% of a typical bacterial genome is noncoding DNA. Infact bacteria use other mechanisms to introduce convenient mutations in their genomes. They can use 3 types of bacterial recombination: conjugation, transformation, and transduction. In bacterial conjugation a special pilus joins the donor and recipient during the DNA transfer. Bacterial transformation consists of the uptake of naked DNA. During Bacterial transduction bacteriophages transfer DNA fragments from one bacterium to another.